\documentclass[a4paper,11pt]{article}
\usepackage{authblk}
\usepackage{hyperref}
\usepackage{color}

\title{Pushing the precision frontier for \\gravitational waves}

\author{Mich\`ele Levi\footnote{\url{https://sites.google.com/view/levim}}}
\affil{Niels Bohr Institute, University of Copenhagen, 
\\Blegdamsvej 17, 2100 Copenhagen, Denmark}
\date{August 6, 2021}

\begin{document}

\maketitle

\begin{abstract}
%Dr Mich\`ele Levi has recently received a prestigious Ernest Rutherford Fellowship from the Science 
%and Technology Facilities Council (STFC). She plans to use this to develop a programme to deliver 
%analytical high-precision predictions for real-world gravitational-wave data and to develop our 
%fundamental understanding of gravity from the smallest quantum to the largest cosmological scales. 
%The programme will address the high demand for accurate theoretical templates of gravitational 
%waveforms from two merging compact objects, such as black holes.

Pushing the precision frontier for gravitational waves is one of the most urgent tasks in 
theoretical physics today, in light of the increasing influx of data from a rapidly growing 
worldwide network of gravitational-wave detectors. Levi will analytically predict gravitational 
radiation from such compact binaries and further develop the use of quantum field theory advances to 
study gravity. She seeks to uncover profound duality relations between gauge and gravity.

International Editor Clifford Holt spoke to Levi about her research, the importance of analytical 
high-precision predictions for real-world gravitational-wave data, and potential challenges as she 
works to develop a better idea of universal commonalities across classical and quantum field 
theories.
\end{abstract}

\clearpage

%\begin{multicols}{2}

\textbf{Firstly, congratulations on being awarded the Ernest Rutherford Fellowship. How important 
are funding mechanisms such as these, and what will the Fellowship mean for your own research and 
career?}

Thank you! A funding mechanism like this prestigious early-career award from the STFC (one of the 
seven UK research councils under UKRI) is very important and unique. This is the most competitive 
scheme in the UK for early-career researchers in my domain, devised to ensure that those 10 
carefully-selected awardees, who are recognised as the very top international talent and leadership 
`material' in science early on in their career, will be integrated into top academic positions in 
the leading research institutions in the UK.

This will enable the awardees to play a key role in the coming decades in the scientific and 
transformative leadership at the national level in the UK and will also represent the UK at the 
international level, solidifying its position as a scientific world-leader for the next generations. 
As a public funder, UKRI selects these future leaders to shape strategic UK policies, to bolster the 
UK's scientific place in the world, and to bring about a radical shift in the current research 
culture in terms of the way that science is conducted, namely its standards and its values, but also 
its role and commitment within society.

Currently, many academics have sadly forgotten that their raison d'etre is to serve the public, that 
they have the privilege of expanding the frontiers of knowledge mainly thanks to the public money 
that is funding their research. We must insist that science that is peer-reviewed and approved for 
publication and funding lives up to the standards of FAIR (Findabilty, Accessibilty, 
Interoperabilty, Reusabilty) principles, namely of responsible research of integrity, that is 
publicly available, widely accessible, and reproducible. We should also advocate for equality, 
diversity, and inclusion in Science, which should properly represent Society, and engage with the 
public. It is very clear that UKRI as a public funder has a strong commitment to realising these 
aspirations, which are unfortunately still overlooked by the majority of private funders.

I am excited and honoured to transfer and base my cutting-edge research programme in the UK, and to 
be part of making the UK a world-leading force in the timely domain of gravitational waves with the 
most innovative theoretical approaches and computational tools applied to gravity. I am also 
incredibly aligned with the aspirations of UKRI to transform the current research culture, and as I 
myself belong to an overlap of under-represented and marginalised groups in science, I am 
particularly aware to how systemic inequalities affect the ability of students and researchers to 
truly realise their potential and excel, or even to just conform to the common skewed metrics of 
academic merit.

I adamantly champion and pursue equality, diversity, and inclusion in every aspect of my everyday 
scientific activities. I believe and I see that it is only through these consistent everyday actions 
and through my being a role model that I can ultimately translate my research work to further 
concrete scientific and societal impact. I also adamantly advance the broad communication of science 
across the research community and through public outreach, which I regard as a social imperative. I 
also find it extremely rewarding to share my knowledge with as many people as possible.

\textbf{Your project aims to deliver analytical high-precision predictions for real world 
gravitational wave data. Can you tell me more about this and why it is important?}

Gravitational radiation is a probing prediction of any complete theory of gravity. Currently, we 
still do not have a complete theory of gravity and this is true for both ends of the scale, since we 
do not know how to describe gravity on quantum scales, which are the smallest scales, and we do not 
have a theory that captures gravity on the largest scales -- cosmological scales, since we do not 
know how to explain the puzzle of dark matter or the accelerated expansion of the Universe that was 
discovered in 1998. Understanding how the force of gravity works on these more `extreme' scales is 
possibly the deepest and most fundamental longstanding open problem in physics.

The first Earth-shaking gravitational-wave (GW) detection announced in 2016 launched a new era of 
gravitational-wave astronomy and precision gravity, which is only in its infancy, with vast future 
impacts in astrophysics and cosmology and, most notably, in fundamental physics. Since this first 
milestone detection, ground-based detectors keep multiplying, and many more are planned for the 
coming years. They are designed to measure various frequency ranges, which will enable a richer 
array of sources and events that generate gravitational radiation to be captured.

The technology of measurements is constantly upgraded, either in currently-operating detectors like 
LIGO or VIRGO, or in upcoming and planned ones, like KAGRA in Japan, and others in the US, Europe, 
India, and Korea. Furthermore, there are also concrete planned space-based experiments from Europe, 
China, and Japan that will also target smaller frequencies and other types of sources. All in all, 
in the coming years and decades we are headed towards a growing influx of gravitational-wave data 
that will have higher and higher quality.

Let me stress what, I believe, much of the community in Europe and the US does not fully recognise 
yet: that Asia is about to emerge as possibly the major player in gravitational-wave astronomy in 
the coming years: in Japan, KAGRA is already operational; in India, IndiGO will become operational 
in just a few years; and in Korea, SOGRO is planned to uniquely target mid-range frequencies from 
Earth, while China and Japan are also planning space-borne detectors, TianQin and DECIGO, 
respectively.

\textbf{How will your research help to inform the activities of detectors such as VIRGO and LIGO, as 
well as perhaps LISA? How could your theoretical work come to complement future 
observations/detections?}

My research generates theoretical data from which a huge bank of templates is created to be matched 
in the detectors by the real data that is measured. The GW signal is weak and is covered by many 
sources of noise so, in order to identify it the matched filtering technique is used for detection, 
which requires the most accurate theoretical signal templates possible. When a match is found, a 
detection is announced. Yet, lots of real-world data, even if is great data is just data without 
interpretation. What transforms data into knowledge is theory and analysis. This is especially true 
for the gravitational-wave signal since it is so weak, so it is really important to separate the 
noise from the true signal.

For the most part, the observable differences among the plethora of candidate theories for gravity 
suggested along the years have been insanely tiny. In most cases, all of these theories seem grossly 
similar within gravitational-wave measurements. So it is only the theoretical resolution into 
various physical effects at a very high-precision, such as those that have been computed in my 
research since my graduate studies, that will enable us to really distinguish among the various 
suggested theories of gravity, and to close in on which is the viable one.

Another feature which is rather unique to my own theoretical framework is that it provides 
predictions for generic sources of compact binaries, not only for binaries of two black holes. As 
the technology in upcoming detectors advances, and as the sensitivity improves, we are able to 
observe more and more neutron stars, for example, in the radiating binaries. Whereas from black 
holes we can `only' learn about gravity theories, neutron stars also tell us about the theory of QCD 
(Quantum Chromodynamics) in extreme conditions that no human-built lab could produce. With future 
detectors, we also expect to observe white dwarfs, and learn more about them.

\textbf{Have you yet been able to identify any potential challenges that you expect to have to 
overcome as you work to develop a better idea of universal commonalities across classical and 
quantum field theories?}

Right now, we see some analogies between interactions of classical extended gravitating objects like 
the huge non-rotating black holes in radiating binaries, and elementary particles of quantum nature, 
which we call `scalar particles' -- these are boson particles of zero quantum spin. But the precise 
mapping between them, and more importantly its generalisations for all types of elementary particle, 
is poorly understood. In particular, my theoretical work pushed the understanding of spinning 
gravitating objects, which would also be analogous to higher-spin particles. Currently, the 
highest-spin particle that has been observed in nature is of spin one, and we also believe there 
could be particles of spin three/two according to supersymmetric theories, as well as our 
theoretically predicted graviton -- a massless particle of spin two. But what about particles with 
spins larger than two? Do they exist? Could they be elementary? Or only show up as what we call 
`composite' particles, which must be composed of elementary ones?

At the moment, we cannot even manage to give well-defined predictions of observables of quantum 
field theories that capture such higher-spin particles, and we are certainly not able to write down 
such quantum field theories explicitly. Furthermore, we are not certain whether this fundamental 
issue exists already at the classical level, i.e.~with the type of two-body interactions that my 
work is tackling for spinning gravitating objects. The big challenge for my research programme, 
therefore, would be to more rigorously and generally link the classical and quantum sides of these 
analogies and to clarify whether there is indeed a fundamental issue with higher-spin -- and, if so, 
whether it originates in the quantisation of the theory, or if it already exists at the classical 
level.

This all relates to possibly the biggest open question in theoretical physics of how to complete the 
theory of gravity at high energies, and therefore it would indeed be a breakthrough for my upcoming 
research programme to be able to provide any further insights on that.

\textbf{What are your short/long term hopes and ambitions?}

Since my graduate studies, I have had the privilege of pioneering and leading innovative analytical 
work and high-precision computation on the highly intricate problem of compact binaries that source 
gravitational radiation. At the time, I did not believe that I would live to see actual detections 
of gravitational waves from these sources, since at that time the detectors had been operational for 
more than a decade without any successful detection.

Yet, the detection came in February 2016, and I can still remember it as though it were yesterday; I 
was overwhelmed with excitement and wonder for what was to come when the first observation of a 
gravitational-wave signal by LIGO was announced that day. Since then, the scientific prospects of 
gravitational-wave science have been continuously improving and exceeding all expectations. The 
future truly looks symphonic, filled with gravitational-wave echoes.

With this increasing influx of data of an ever improving quality, we have a chance to really push 
theoretical predictions and to test so much more theory in gravity and even in QCD. My research 
programme is exactly aimed at maximising the potential of what we can learn from gravitational-wave 
measurements on the theory of gravity across all scales.

I hope to engage the broader scientific community with this exciting new domain of research, such 
that it really crosses over all sub-fields of theoretical physics, especially since this is a great 
scientific challenge that can benefit from so many different specific perspectives. I believe that 
every one of these perspectives can bring in something fresh and interesting to the mix. I also 
believe that, to some extent, I have already contributed to the widespread attention that this type 
of research obtained in the traditional high energy physics community. In the UK, I hope to engage 
more people and institutions with the research in this field as the UK really has the potential to 
become a superpower in this domain, especially in the type of analytical work that I am involved in, 
which has been traditionally dominated by France.

I am also a big believer in open code and open data, and I really want to see individual 
researchers, big experimental collaborations, scientific journals, and academic institutions all 
persistently moving towards this in the coming years. The scientific challenges at hand are too big, 
and we need to focus on the science and make our best joint efforts to make breakthroughs, instead 
of yielding to the old zero-sum game mentality that keeps theoretical physics stagnated.

Simply put, I just hope that, together, we can all increase our knowledge of the world around us. 
Isn't that the greatest privilege of our society?

%\newpage
\section*{Acknowledgements}

%This is to appear only in the arXiv preprint.}
\noindent ML received funding from the European Union's Horizon 2020 research and 
innovation programme under the Marie Sk{\l}odowska-Curie grant 847523.  
\\Her upcoming research program will be supported by the STFC Rutherford grant ST/V003895/1 
\textit{``Harnessing QFT for Gravity''}.

%\end{multicols}

\end{document}